\documentclass[twocolumn,groupedaddress,aps,prl,amsmath,amssymb, preprintnumbers, superscriptaddress,showpacs]{revtex4-2}

\usepackage{newtxtext, newtxmath}

\usepackage{color, graphicx}     
\usepackage{dcolumn}     
\usepackage{bm}                  
\usepackage{amssymb}
\usepackage{latexsym}
\usepackage{amsfonts}
\usepackage{amsmath}
\usepackage{multirow}

\usepackage[colorlinks=true,citecolor=blue]{hyperref}
\hypersetup{colorlinks=true,citecolor=blue,linkcolor=red,urlcolor=blue}

\usepackage[dvipsnames]{xcolor}
\usepackage{times}
\begin{document}


\makeatletter
\newcommand*\bigcdot{\mathpalette\bigcdot@{.3}}
\newcommand*\bigcdot@[2]{\mathbin{\vcenter{\hbox{\scalebox{#2}{$\m@th#1\bullet$}}}}}
\makeatother

\title
{Unconventional Thermophotonic Charge Density Wave}

\newcommand*{\IUF}[0]{{Institut Universitaire de France, 1 rue Descartes, F-75231 Paris, France}}

\newcommand*{\NUS}[0]{{Department of Electrical and Computer Engineering, National University of Singapore, Singapore 117583, Singapore}}

\newcommand*{\HIT}[0]{{School of Energy Science and Engineering, Harbin Institute of Technology, Harbin 150001, China}}

\newcommand*{\IPM}[0]{{School of Nano Science, Institute for Research in Fundamental Sciences (IPM), 19395-5531 Tehran, Iran}}

\newcommand*{\DIPC}[0]{{Donostia International Physics Center (DIPC), 20018 Donostia-San Sebasti\'an, Spain}}

\newcommand*{\IFS}[0]{{Centre for Advanced Laser Techniques, Institute of Physics, 10000 Zagreb, Croatia}}

\newcommand*{\LCC}[0]{{Laboratoire Charles Coulomb (L2C), UMR 5221 CNRS-Université de Montpellier, F-34095 Montpellier, France}}

\author{Cheng-Long Zhou}
\thanks{These two authors contributed equally}
\affiliation{\HIT}%
\affiliation{\NUS}%
\author{Zahra Torbatian}
\thanks{These two authors contributed equally}
\affiliation{\IPM}%
\author{Shui-Hua Yang}
\affiliation{\NUS}%
\author{Yong Zhang}
\affiliation{\HIT}
\author{Hong-Liang Yi}
\email{Corresponding author: yihongliang@hit.edu.cn}
\affiliation{\HIT}%
\author{Mauro Antezza}
\affiliation{\IUF}
\affiliation{\LCC}%
\author{Dino Novko}
\email{Corresponding author: dino.novko@gmail.com}
\affiliation{\IFS}
\affiliation{\DIPC}
\author{Cheng-Wei Qiu}
\email{Corresponding author: chengwei.qiu@nus.edu.sg}
\affiliation{\NUS}

\begin{abstract}
Charge-order states of broken symmetry, such as charge density wave (CDW), are able to induce exceptional physical properties, however, the precise understanding of the underlying physics is still elusive. Here, we combine fluctuational electrodynamics and density functional theory to reveal an unconventional thermophotonic effect in CDW-bearing TiSe$_2$, referred to as thermophotonic-CDW (\textit{tp}-CDW). The interplay of plasmon polariton and CDW electron excitations give rise to an anomalous negative temperature dependency in thermal photons transport, offering an intuitive fingerprint for a transformation of the electron order. Additionally, the demonstrated nontrivial features of \textit{tp}-CDW transition hold promise for a controllable manipulation of heat flow, which could be extensively utilized in various fields such as thermal science and electron dynamics, as well as in next-generation energy devices.
\end{abstract}
\maketitle
 Strongly correlated quantum systems can form intriguing fundamental collective modes of broken symmetry, such as charge density waves (CDW), characterized with a modulation of the valence electron density and a corresponding lattice distortions\,{\cite{gruner1988dynamics,rossnagel2011origin}}. Microscopic origins of these charge-order states are still actively debated, especially for two-dimensional and bulk systems, where proposed mechanisms\,{\cite{rossnagel2011origin,zhu15}} range from purely electron-\,{\cite{perdew21}} to purely lattice-driven process\,{\cite{zhou20}}. For instance, the nature of CDW in transition metal dichalcogenides (TMDs), such as $1T$-TiSe$_2$, is widely discussed in terms of excitonic insulator scenario\,{\cite{disalvo76,cercellier07}}, aniostropic electron-phonon mechanism\,{\cite{watson19}}, anharmonicity\,{\cite{zhou20,zheng22}}, or Mott-related electronic correlations\,{\cite{raghu08}}. In order to reach a consensus it is important to achieve a microscopic visualization of CDW states by using various complementary scattering techniques and directly determine and comprehend the corresponding fingerprints of structural and Fermi surface modifications, such as periodic lattice distortions and CDW gap openings\,{\cite{yang2022visualization,nguyen2023ultrafast,hildebrand18,watson19}}.
 \setlength{\parskip}{0pt}
 
 The CDW states was found to induce some remarkable electrodynamical features that are highly tunable with temperature \,{\cite{schwartz1995fluctuation}} and could be utilized in switchable optoelectronic devices, such as unconventional resistivity change\,{\cite{disalvo76}}, anomalous plasmon renormalization\,{\cite{lin22}} and hybridization \,{\cite{song21,torbatian23}}, as well as second-harmonic generation\,{\cite{zhang22}}. Furthermore, thermophotonics is ubiquitous in nature, from gigantic galaxies to microscopic atomic structures\,{\cite{zhang2007nano}}. According to earlier studies, there is a preference to concentrate on the energy characteristics of thermal photons, for the purpose of enhancing the energy intensity of thermal photons or developing novel manipulation techniques\,{\cite{biehs2021near,xu2021broadband,liu2022thermal}}. However, this energy characteristics of thermal photons may also hold the key to our understanding of new physics, as in the case of the light quantum hypothesis hidden in black body radiation. It is worth pondering whether there is an apparent energy fingerprint of thermal photons for the CDW transition that can be used to identify its features and provide valuable information on these collective modes. Understanding the thermophotonic properties of CDWs could also help  advance our knowledge of how electron order states and photons interact in condensed matter systems and improve the efficiency of various thermal technologies. 
 \setlength{\parskip}{0pt}
\begin{figure*}
	\centering
	\centerline{\includegraphics[width=0.98\textwidth]{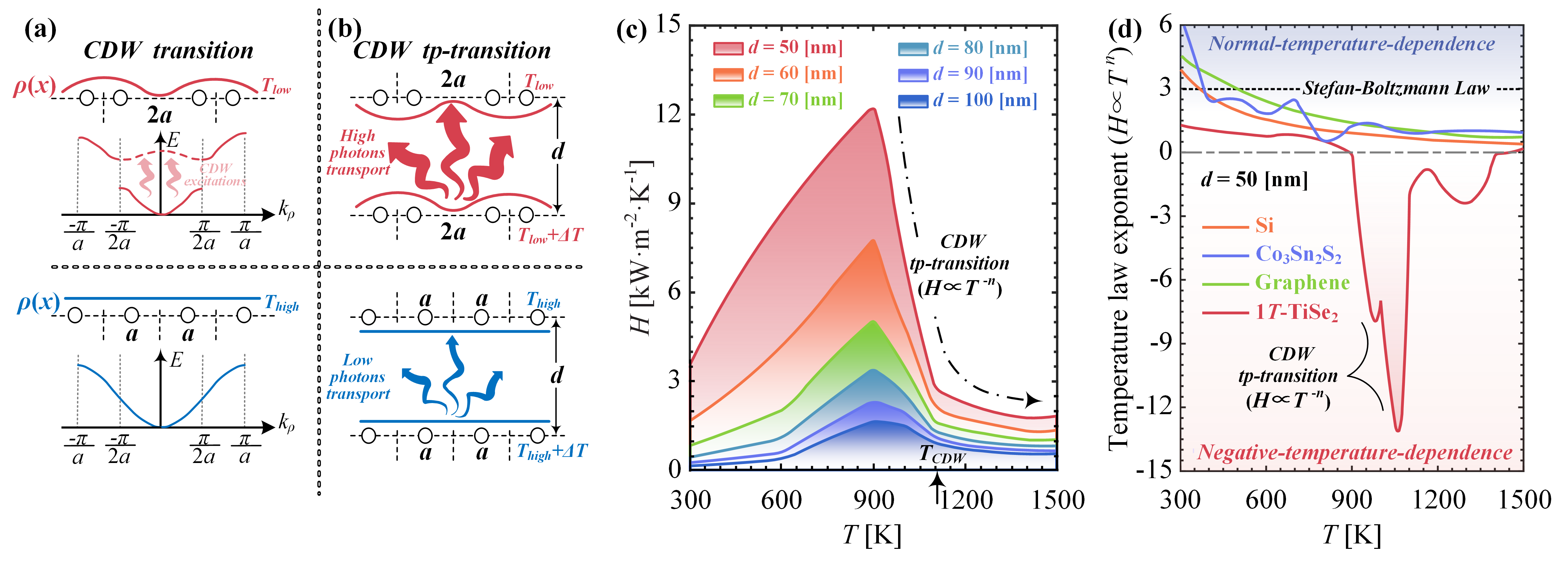}}
	\caption{(a) Schematics of the structure and electron dispersion before (after) the CDW transition, including the CDW-induced interband electron excitations. (b) Schematics of thermophotonic CDW transition for the CDW (top) and standard unit cells (bottom). The thermal photons transport between two CDW-bearing materials with vacuum gap \textit{d}. (c) The calculated energy transport coefficient $H$ of single-layer TiSe$_2$ for various vacuum gaps as a function of $T$. (d) The exponent of the temperature power law as a function of temperature for CDW-bearing TiSe$_2$ and other representative plasmonic materials (such as graphene\,{\cite{ilic2012near}}, Si\,{\cite{desutter2019near}}, and Weyl semi-metal Co$_3$Sn$_2$S$_2$\,{\cite{kotov2018giant}}) for $d = 50$\,nm. A positive exponent means that the thermophotonic intensity increases with temperature , while a negative exponent means the opposite. The black dotted line represents Stephen Boltzmann's law of black body ($H_{bb}=4 \sigma T^3$).}
	\label{Fig1}
\end{figure*}

 In this Letter, we reveal the influence of CDW on the thermophotonic behavior for the TiSe$_2$ single layer,  which we refer to as thermophotonic CDW transition (\textit{tp}-CDW transition), by means of fluctuational electrodynamics (FED)\,{\cite{ilic2012near,carminati1999near}}, current-current linear response\,{\cite{novko16}}, and density functional theory (DFT), connecting two rapidly growing areas of physics research, namely, CDWs and thermophotonics. The results show that during \textit{tp}-CDW transition, photonic energy transport produces a significant negative temperature dependence, i.e., $H\propto T^{-n}$, violating the Stefan-Boltzmann prediction $H \propto T^{3}$\,{\cite{volokitin2007near}}. This negative temperature dependence is highly non-trivial and suggests an intimate connection between the CDW order and thermophotonics. We then explore the underlying physical mechanisms of nontrivial signatures during \textit{tp}-CDW transition and investigate the possible effects of an intricate hybrid mode consisting of plasmons and CDW excitations in thermal photons transport. Moreover, the present results suggest that \textit{tp}-CDW transition would be a promising energy transfer platforms with unconventional features.

Let us consider a system composed of two CDW-bearing materials brought into proximity with a vacuum gap size of \textit{d}, as sketched in Fig.\,\ref{Fig1}.  In this work, our aim is to investigate the thermophotonic signatures of the CDW-bearing material. To this end, we need to make sure that the system is a strongly coupled system, i.e., that the emitter and receiver have the same electromagnetic resonance nature. Hence, we assume the temperature of the source to be $T+$$\Delta$$T$, with $T$ the temperature of the receiver. Below the CDW transition temperature $T_{\rm CDW}$, the lattice distortion due to the electron-lattice interaction induces a periodic modulation of the electron density $\rho$ [Fig.\,\ref{Fig1}(a)], opening a CDW electronic gap, and allowing a strong thermal photon transport channel. With increasing temperature, the CDW unit cell $2a$ with periodic lattice distortions is modified to a standard $1a$ unit cell, in parallel with the disappearance of the CDW order [Fig.\,\ref{Fig1}(b)]. In our work we consider peculiar quasi-two-dimensional (Q2D) TiSe$_2$, hosting a CDW phase with the $2\times 2$ distorted structure\,{\cite{rossnagel2011origin,disalvo76}}. 
 
Our calculation of the thermal photons transport follows the standard FED procedure. The fluctuating in-plane surface currents $\bar{\textbf{\textit{j}}}$ (\textbf{\textit{r}}, \textit{t}) in each sheet obey the fluctuation dissipation theorem\,{\cite{ilic2012near,rincon2022enhancement}} 
\begin{align}
  <\bar{\textbf{\textit{j}}}_l(\textbf{\textit{r}}, \textit{t})\bar{\textbf{\textit{j}}}_m(\textbf{\textit{r'}}, \textit{t'})>=&\delta_{lm}\int\frac{d^2k_{\rho}d\omega}{(2\pi)^3} \hbar\omega {\rm coth}\frac{\hbar\omega}{2T}\nonumber\\
  & \times Re(\frac{-i}{\omega} {\rm lim}_{k_{\rho  \rightarrow 0}} \Pi(k_{\rho}, \omega, T))\nonumber\\
  & \times e^{ik_{\rho}(\textbf{\textit{r}}-\textbf{\textit{r'}})-i\omega(\textit{t}-\textit{t'})},\label{eq:1}
\end{align}
where $k_{\rho}$ is the in-plane two-dimensional wave vector and
\textit{l}(\textit{m}) labels the orthogonal in-plane directions. The fluctuating electric fields $\textbf{\textit{E}}(\textbf{r}, t)$ can be derived from the thermal random current $\bar{\textbf{\textit{j}}}$ in Eq.\,\eqref{eq:1} by the Green's tensors \cite{Latella2015near}. The average Joule loss power, and therefore the energy transfer coefficient between the emitter and the receiver, can then be calculated from the electric field and the electric current \cite{tang2021near}
\begin{align}
\label{eq:2}
H(T)&=\frac{<\bar{\textbf{\textit{j}}}^E\cdot\textbf{\textit{E}}^E>_{T+\Delta T}-<\bar{\textbf{\textit{j}}}^R\cdot\textbf{\textit{E}}^R>_{T}}{\Delta T}\nonumber\\
& =\int d\omega \frac{\partial\Theta(\omega, T)}{\partial T}f(\omega, T).
\end{align}
Here
\begin{align}
\label{eq:3}
f(\omega, T)=\int \frac{\xi k_{\rho}d k_{\rho}}{(2\pi)^2},
\end{align}
where $\xi=\underset{j=p,s}{\Sigma}\chi_{E}\chi_{R}|e^{ik_{z}d}|^{2}/|1-r_{E}r_{R}e^{2ik_{z}d}|$. $\Theta(\omega, T) =\hbar\omega/(e^{\hbar\omega/k_{b}T}-1)$ is the average energy of a photon at frequency $\omega$, and $k_{b}$ is the Boltzmann constant. $f(\omega, T)$ is the spectral transfer function, characterizing frequency dependence of the energy transport. The $r$ and $\chi$ are reflection and emissivity coefficients, respectively (see Supplemental Material (SM)\,{\cite{SM}} for explicit and rather standard expressions). $\Pi(k_{\rho}, \omega, T)$ is noninteracting current-current response tensor \,{\cite{novko16}}, which embodies the charge density correlation that describes the response of CDW-bearing material in terms of induced charge density to the external potential.

This charge density correlation is calculated by using the plane-wave density-functional-theory code \textsc{Quantum Espresso}\,{\cite{giannozzi2009quantum}} and the semi-local exchange-correlation Perdew-Burke-Ernzerhof (PBE) functional. The phonon dynamics and electron-phonon coupling, which are found to be critical to the CDW properties in TiSe$_2$\,{\cite{calandra2011charge,novko22}}, are obtained within the density functional perturbation theory framework\,{\cite{dfpt}}, and then interpolated with maximally-localized Wannier functions\,{\cite{marzari12,mostofi2008wannier90}} and EPW code\,{\cite{epw23}}. Further computational details can be found in the SM\,{\cite{SM}}.

 In order to determine the temperature dependence, the energy intensity of thermal photons can be fitted to an analytical expression similar to Stefan-Boltzmann law\,{\cite{lucchesi2021temperature}}
 \begin{align}
H \propto C \sigma T^{n},
\end{align}
with $\sigma$ being the Stefan-Boltzmann constant and $n$ the temperature power-law index. The fitting parameter is $C$, a prefactor related to the vacuum gap and temperature. Figure\,\ref{Fig1}(c) shows the energy transport coefficient of TiSe$_2$ as a function of $T$ and various vacuum gaps $d$. For the Stefan-Boltzmann law ($H_{bb}=4 \sigma T^3$), the temperature power is strictly equal to 3 {\cite{zhang2020optimal}}. Although other plasmonic materials (such as graphene\,{\cite{ilic2012near}}, Weyl semimetals\,{\cite{kotov2018giant}}, and Si\,{\cite{desutter2019near}}) also significantly deviate from the Stefan–Boltzmann prediction ($n\neq3$), they still follow a positive temperature dependence [see Fig.\,\ref{Fig1}(d)]. When the lattice is stable ($T \gg T_{\rm CDW}$ or $T \ll T_{\rm CDW}$), the energy transport coefficient of TiSe$_2$ also abides by the above positive trend. However, when the temperature increases towards $T_{\rm CDW}$ the results for 1\textit{T}-TiSe$_2$ challenge this standard perspective, showing the counter-intuitive phenomenon of thermal photons with a negative temperature dependence. Intriguingly, this anomalous drop is exactly the opposite of the enhancement in the phonon thermal conductivity induced by CDW order change\,{\cite{kwok1989thermal}}. This negative temperature dependence is entirely caused by a change in the electrodynamical properties as a result of the \textit{tp}-CDW transition, which is in contrast with the previous phenomenon of negative differential thermal conductance due to the temperature-related decoupling of the emitter and receiver\,{\cite{zhu2012negative,feng23}}. 
 
 This trend continues to well above the CDW phase change and peaks around $T_{\rm CDW}$. Since the present results are obtained with the PBE exchange-correlation DFT functional the transition temperature $T_{\rm CDW}$=1105\,K is overestimated, while more accurate result could be obtained with a proper inclusion of electron correlations and anharmonic effects\,\cite{zhou20}. Despite that, the closing of the gap and the relative difference between the $T_{low} (\ll T_{\rm CDW})$ and $T_{high} (\ge T_{\rm CDW})$ are in a good agreement with the experimental results as obtained with angle-resolved photoemission spectroscopy \,{\cite{li07}} and resonant inelastic x-ray scattering \,{\cite{monney2012mapping}}. Therefore, with the inclusion of phonon-phonon correction we only expect a correction to the value of $T_{\rm CDW}$, while anharmonic effects should have no impact on optical and thermophotonic properties reported in the present work (see Sec. S5 in Ref.\,{\cite{SM}}).
 
\begin{figure}
 	\centering
 	\centerline{\includegraphics[width=0.98\columnwidth]{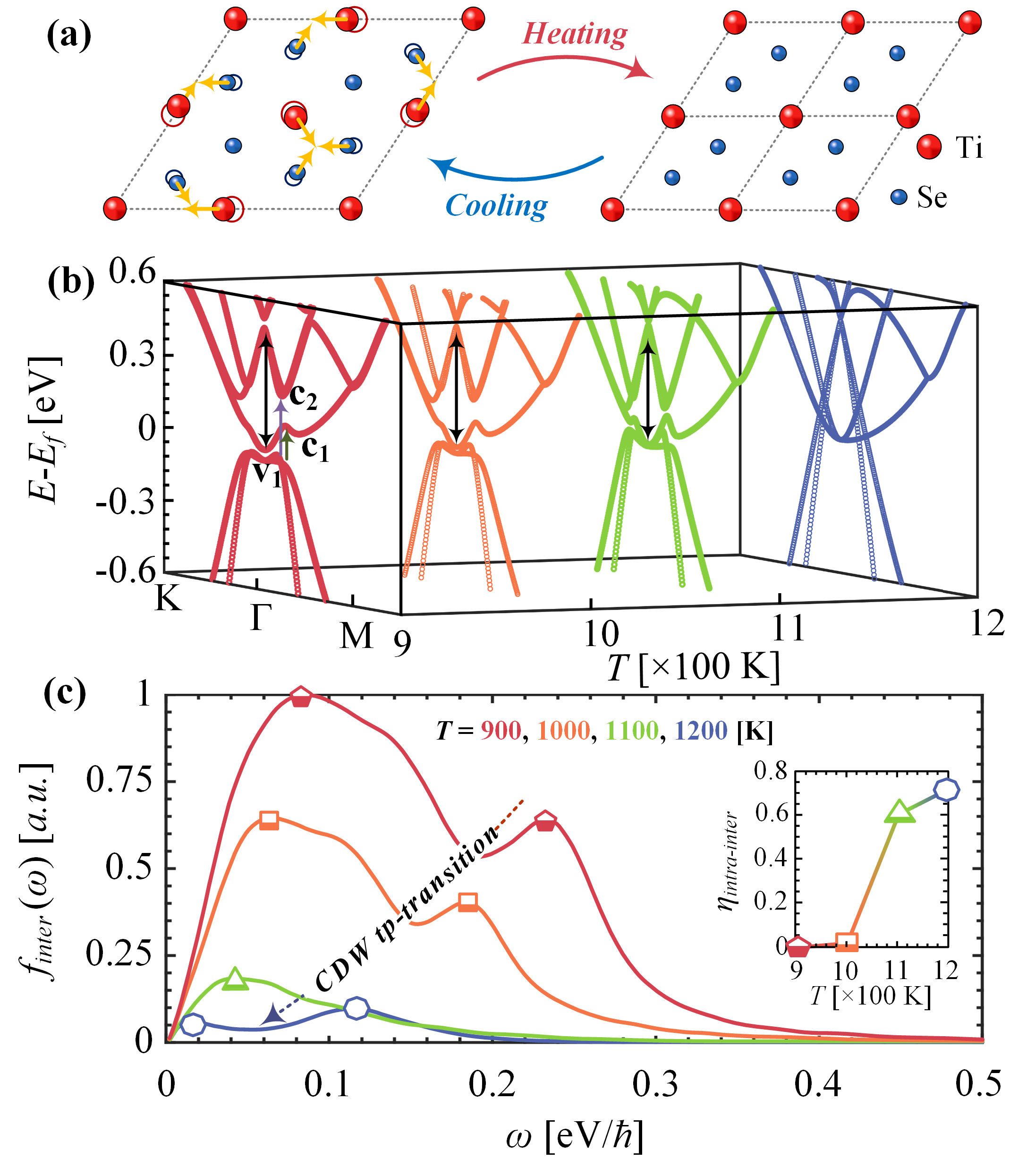}}
 	\caption{(a) The CDW and standard crystal structures of TiSe$_2$. (b) The calculated electronic band structure of TiSe$_2$ along high-symmetry points of the Brillouin zone at different temperatures. (c) The interband spectral intensity function $f_{inter}$($\omega$) at different temperatures for $d=50$\,nm. The inset gives the spectral intensity ratio $\eta_{inter}^{intra}=f_{intra}/f_{inter}$ between interband and intraband processes.}
 	\label{Fig2}
 \end{figure}

 Furthermore, results show that the energy transport of thermal photons decays directly before $T_{\rm CDW}$ as the temperature increases. This sharp evolution ahead of $T_{\rm CDW}$ is also found in the resistivity behavior of TiSe$_2$\,{\cite{disalvo76,knowles2020fermi}}. The strong temperature dependence below $T_{\rm CDW}$ is due to a continuous reconstruction of the electronic band structure as the temperature increases [see  Figs.\,\ref{Fig2}(a)  and \ref{Fig2}(b)], with a continuous decline in accordance with the mean-field theory result for a second-order transition {\cite{gruner2017charge}}. This is significantly different from the rapid transformation of VO$_2$ from a monoclinic to a rutile crystal structure, which is a first-order transition {\cite{lee2017anomalously}}. 

Figure\,\ref{Fig2}(c) demonstrates the modifications of the spectral intensity function induced by interband electronic excitations $f_{inter}$ across the \textit{tp}-CDW transition.  This spectral intensity function represents the energy levels carried by thermal photons of different frequencies (see also Ref.\,{\cite{SM}} for the experssions).  We observe a sharp drop in $f_{inter}$ when the temperature is increasing towards $T_{\rm CDW}$, which is responsible for the negative temperature dependence of the thermal photons as shown in Fig.\,\ref{Fig1}. The inset in Fig.\,\ref{Fig2}(c) shows the spectral intensity ratio between the intraband and interband processes $\eta_{inter}^{intra}=f_{intra}/f_{inter}$. It can be seen that the intraband contributions are negligible below $T_{\rm CDW}$, while interband and intraband contributions progressively approach the same level for $T \ge T_{\rm CDW}$. Overall, the powerful thermal photons transport across the \textit{tp}-CDW transition is dominated primarily by the interband processes. Figures\,\ref{Fig2}(a) and \ref{Fig2}(b) show the unit cells and the electronic band structure of TiSe$_2$ to visualize the CDW order change. As the CDW order change occurs, the commensurate $2\times2$ structure with periodic lattice distortions is modified into a standard $1\times 1$ unit cell. As shown in Fig.\,\ref{Fig2}(b), this structural transition is accompanied by visible  modifications of the electronic band structure. At low temperatures, $T<T_{\rm CDW}$, the CDW order leads to the opening of the CDW gap between the Se-4$p$ valence states and the Ti-3$d$ conduction bands (double-sided black arrow), which opens the possibility for interband electron transitions (one-sided colored arrows), i.e., interband channels of thermal photons. The results show the contributions from the two interband transitions, leading to pronounced corresponding peaks in $f_{inter}$, where the high-energy peak comes from excitations between the highest valence band $v_1$ and the second lowest conduction band $c_2$ (purple arrow) , while the low-energy contribution is due to transitions between $v_1$ and lowest conduction band $c_1$ (green arrow).
As the temperature increases, the CDW gap gradually decreases, leading to the suppression of the CDW interband processes. When the temperature exceeds $T_{\rm CDW}$, the closing of the CDW gap is responsible for the significant attenuation of $f_{inter}$. The two pronounced excitation peaks are closely related to the excitation peaks as observed in infrared spectroscopy\,{\cite{li07}} and resonant inelastic x-ray scattering\,{\cite{monney2012mapping}}.
 \begin{figure}
 	\centering
 	\centerline{\includegraphics[width=0.98\columnwidth]{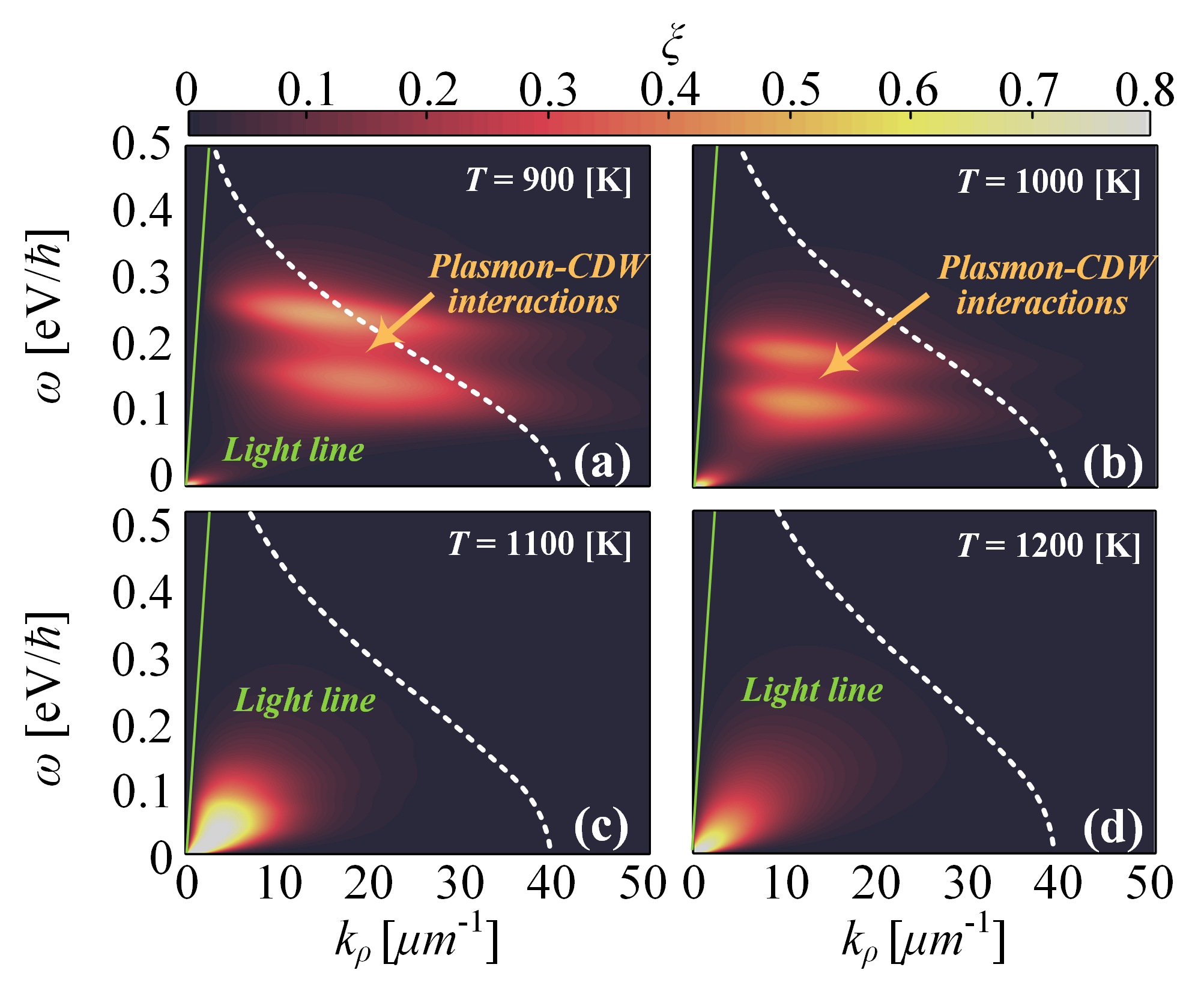}}
 	\caption{The evolution of thermophotonic probability during the \textit{tp}-CDW transition for (a) $T = 900$\,K, (b) $T=1000$\,K, (c) $T=1100$\,K, and (d) $T=1200$\,K. The green line is dispersion relation of light in a vacuum. The white dotted lines in all panels correspond to the occupation factor $\partial\Theta(\omega;T)/\partial T$ in arbitrary units, where  $\Theta(\omega;T)=\hbar\omega/\exp(\hbar\omega/k_bT)$.}
 	\label{Fig3}
 \end{figure}
 \setlength{\parskip}{0pt}

To further clarify this nontrivial thermophotonic behavior, we show the thermophotonic probability (see also Ref.\,{\cite{SM}}).  The thermophotonic probability represents the tunneling probability of thermal photons between the emitter and receiver\,{\cite{salihoglu2023nonlocal,song2016radiative}}. In our work, the polarization states are all transverse-magnetic waves {\cite{SM}}. The dependence of the thermophotonic  probability on the CDW order is shown in Fig.\,\ref{Fig3}. For the cases of $T = 900$\,K and 1000\,K, thanks to the strong interband excitation provided by the CDW gap, the bright band of thermophotonic probability is diffused into a wider region in momentum space. Interestingly, similar to the plasmon-exciton interactions {\cite{pincelli2023observation}}, the CDW excitations appearing below the $T_{\rm CDW}$ can also couple to the Q2D plasmon, forming a hybrid CDW-plasmon mode\,{\cite{song21,torbatian23}}, and leading to an anti-crossing phenomenon of the thermophotonic probability (see also Ref.\,{\cite{SM}} for the dispersion).
 
 As the temperature increases ($T \ge T_{CDW}$), we find that the bright band of thermophotonic probability is compressed into a smaller region in momentum space until the CDW gap is closed. For the cases of $T = 1100$\,K and 1200\,K, the photonic energy with high-$k_{\rho}$ and high-$\omega$ is poorly excited by localized emitters, as seen in Figs.\,\ref{Fig3}(c) and \ref{Fig3}(d), which is consistent with our above observations for the spectral result. On the other hand, when the CDW disappears, a strong thermophotonic excitation band appears in the region with low-$k_{\rho}$ and low-$\omega$. The reason for this is that when the CDW gap vanishes, the plasmon couples with the phonons (in particular, soft CDW phonons), achieving a plasmon-phonon scattering effect; this plasmon-phonon scattering channel can lead to a sharp increase in the plasmon decay rate and hence in the intraband plasmon energy. Although the transmission maxima above $T_{\rm CDW}$ are higher than that induced by the CDW order below $T_{\rm CDW}$ [Fig. \ref{Fig3}(c)], the wavevector and bandwidth are too small. It can be seen that this  small range of wavevectors significantly weakens the performance of thermal photons, which in turn induces the negative temperature dependence of thermophotonic intensity.
 \begin{figure}
 	\centering
 	\centerline{\includegraphics[width=0.98\columnwidth]{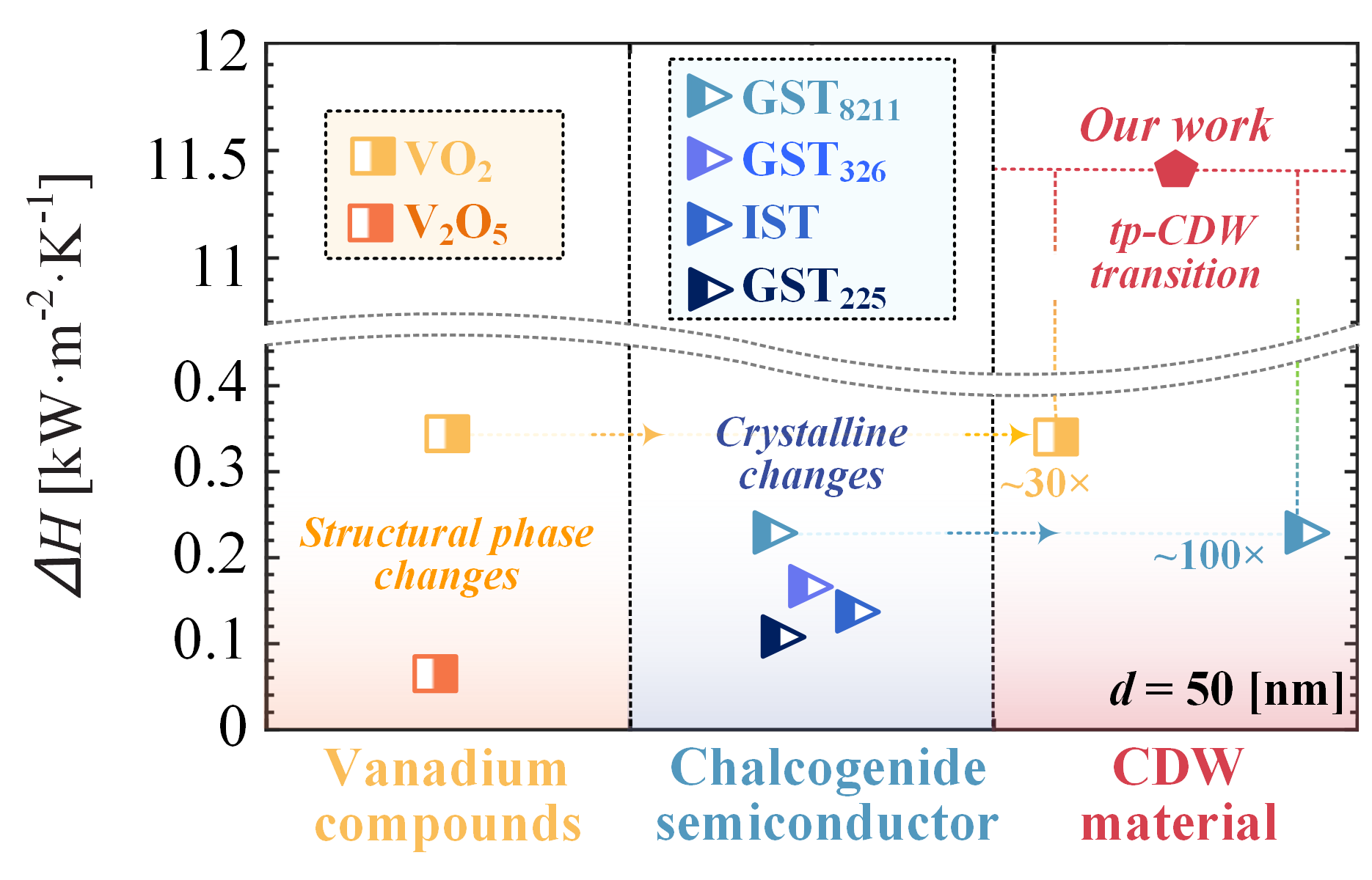}}
 	\caption{A comparison of the thermal management coefficient $\Delta H$ of the present CDW material and different materials reported in literature\,\cite{Latella2021near}. Thermal management coefficient $\Delta H$ is defined as the difference between the highest and the lowest energy transport coefficient during the phase change process.}
 	\label{Fig4}
 \end{figure}
 \setlength{\parskip}{0pt}
 
 As with the current control, the flexible management of heat flow has always been a sought-after goal and is of great importance in many areas of physics and chemistry {\cite{li2012colloquium,Latella2021near}}. Analogous to the modern electronics, a number of temperature-controlled thermal components for controlling the heat flow has been proposed {\cite{Liu2020near}}, such as thermal switch {\cite{van2012tuning,li2019adiabatic}}, thermal diodes {\cite{otey2010thermal}}, thermal memory {\cite{kubytskyi2014radiative}}, and thermal transistor {\cite{ben2014near}}. The \textit{tp}-CDW transition discussed above seem to be an excellent tool for realizing these thermal analogs of modern electronics. 
 It is also important to note that the interband transitions can also be modulated by the external field to achieve artificial energy tunability \cite{He2020near,Li2021NRM,Messina2012near}. Unfortunately, such robust interband transitions may not be readily useful for temperature-controlled thermal devices and we need to discover a mechanism for interband transitions that can be affected by temperature and that allows for high-efficiency control of the heat flow. Here, to categorize the temperature-controlled thermal management potential of materials, we introduce the thermal management coefficient $\Delta H$, that is, the difference between the highest and the lowest energy transport coefficient during the phase change process. Figure \ref{Fig4} compares the $\Delta H$ of several other reported temperature-controlled thermal management materials (vanadium compounds and chalcogenide semiconductors). The vacuum gap is fixed at 50 nm.
 This vanadium compounds (such as VO$_2$ {\cite{tang2020thermal}} and V$_2$O$_5$ {\cite{littlejohn2017naturally}}) and the chalcogenide semiconductors (such as IST {\cite{hessler2021in3sbte2}} and GST {\cite{sreekanth2018ge2sb2te5}}) use lattice phase transition and lattice disorder alteration, respectively, to achieve temperature-controlled thermal management. Note that the $\Delta H$ values of the vanadium and the chalcogenide semiconductor compounds are lower than those predicted here for the CDW-bearing TiSe$_2$. Compared to these two mechanisms, thanks to its second-order transitions features, the \textit{tp}-CDW transition could highlight a wider temperature manipulation range, thus providing a stronger thermal operating capacity and a larger design freedom for thermal devices. 
  \setlength{\parskip}{0pt}
 
 In conclusion, we have constructed a connection between the underlying electron order state and thermophotonics. The results show a nontrivial signature of thermal photons in the CDW material, in which thermophotonic intensity presents a negative temperature dependence, in contrast to the predictions of the Stefan-Boltzmann law. We have further shown that the nontrivial signature naturally originates from the suppression of the interband excitation associated with the annihilation of the CDW electronic band gap. The combination of the CDW and thermophotonics provides an alternative probing mechanism for studying the evolution of the electron order states, shedding light on microscopic origin of the CDW phase. It would be interesting to extend this work to study other quantum phases in two dimensions, such as Mott insulated and Wigner crystal states. Besides, the anomalous thermal response of the CDW phase also hints at its fascinating potential in heat flow manipulation, which opens up new possibilities for thermal physics and thermal applications.
  \setlength{\parskip}{10pt}
 
 \begin{acknowledgments}
C.W.Q. would like to acknowledge the support from Ministry of Education, Singapore, via the grant A-8000107-01-00. This research / project is also partially supported by the National Research Foundation, Singapore (NRF) under NRF’s Medium Sized Centre: Singapore Hybrid-Integrated Next-Generation Electronics (SHINE) Centre funding programme. H.-L. Y. acknowledges the support from the National Natural Science Foundation of China (Grant No. U22A20210). Y. Z. acknowledges the support from the National Natural Science Foundation of China (Grant No. 52106083). C.-L. Z. acknowledges the support from the China Scholarship Council (CSC) (Grant No. 02206120132). D.N. acknowledges financial support from the Croatian Science Foundation (Grant no. UIP-2019-04-6869). M.A. acknowledges the grant ”CAT”, No. A-HKUST604/20, from the ANR/RGC Joint Research Scheme sponsored by the French National Research Agency (ANR) and the Research Grants Council (RGC) of the Hong Kong Special Administrative Region. Part of the computational resources were provided by the DIPC computing center.
 \end{acknowledgments}

\bibliography{reference1}

\end{document}